\documentstyle[aps,epsf,psfig]{revtex}

\begin{document}
\twocolumn

\title{Break junctions of the heavy-fermion superconductors}
        
\author{K.~Gloos$^{a}$, F.~B.~Anders$^{a,b}$, 
B.~Buschinger$^{a}$, and C.~Geibel$^{a}$}
\address{$^a$ Institut f\"ur Festk\"orperphysik,
 Technische Hochschule Darmstadt, D-64289 Darmstadt, Germany}
\address{$^b$ Department of Physics, The Ohio State University, 
 Columbus, Ohio, 43210-1106, USA}

\date{August, 19. 1996}

\maketitle

\begin{abstract}
Mechanical-controllable break junctions of the heavy-fermion superconductors
can show Josephson-like superconducting anomalies. But a systematic 
study on the contact size demonstrates that these anomalies are 
mainly due to Maxwell's resistance being suppressed in the superconducting 
heavy-fermion phase. Up to day, we could not find any superconducting 
features by vacuum-tunnelling spectroscopy, providing further evidence for 
the pair-breaking effect of the heavy-fermion interfaces.
\end{abstract}

\pacs{\\
 KEYWORDS: \\[-0.3cm]
 Break junctions, heavy-fermion superconductors, Andreev reflection, 
 \\[-0.3cm] Josephson effect \\
 \\
 CORRESPONDING AUTHOR \\[-0.3cm]
 Dr. K. Gloos \\[-0.4cm]
 Institut f\"ur Festk\"orperphysik \\[-0.4cm]
 Technische Hochschule Darmstadt \\[-0.4cm]
 Hochschulstr. 8 \\[-0.4cm]
 D - 64289 Darmstadt \\[-0.4cm]
 GERMANY \\
 FAX:    49 - 61 51 - 16 48 83 \\[-0.4cm]
 EMAIL:  DF15@HRZPUB.TH-DARMSTADT.DE
}


Point contacts with heavy-fermion (HF) superconductors (SC) have attracted 
much interest during the past years. The finite contact resistance due to 
the small contact area allows charge carriers to be accelerated across the 
interface, and thus to observe the energy-dependent scattering processes. 
Andreev reflection (AR) and Josephson effect (JE) promise direct access to 
the still unknown SC order parameter of these compounds.

The most spectacular results have been obtained with SC counter-electrodes.
Poppe et al. \cite{Poppe85}, Han et al. \cite{Han86} as well as Nowack et 
al. \cite{Nowack95} found JE-like anomalies with HFSC in contact with 
Al, Ta, Mo, Nb, and NbTi. 
He et al. \cite{He92} verified that HFSC is really based on Cooper 
pairs, using a UPd$_2$Al$_3$ sample as part of a SQUID. 
On the other hand, Moreland et al. \cite{Moreland94} could find 
neither supercurrents on metallic contacts nor SC features by 
vacuum-tunnelling spectroscopy on UBe$_{13}$ break junctions. This was 
attributed to the pair-breaking effect of the surface or interface.

Until recently, the SC anomalies of point contacts with HFSC and normal 
metals have been interpreted as clear-cut signature of AR, see e.g. 
\cite{Nowack87,Hasselbach92,Nowack92,Goll93,Dewilde94,Samuely95,Naidyuk96}. 
However, the correct interpretation has to take into account the high 
specific resistivity $\rho (T)$ these HFSC have in the normal-state. The 
SC anomalies usually observed turned out to be essentially due to normal 
backscattering, thermal, and diffusive processes, while the AR signal 
can hardly be resolved \cite{Gloos96-ube13,Gloos96-icps2,Gloos96-scal}. 
This missing AR signal may be explained by the short lifetime of the heavy 
quasi-particles or by a normal interface layer formed on preparing the 
contacts \cite{Gloos96-scal,Anders-sces}. Thus the 'quality' of the 
interface is a very important parameter. Mechanical-controllable break 
junctions seem to be the most effective method to get clean interfaces, 
not degraded by oxide layers, if they are prepared in situ at low 
temperatures and at UHV conditions. 

To analyse those junctions one must distinguish the different processes 
that contribute to the contact resistance $R$, and also to the 
SC anomalies. For details see Ref. \cite{Gloos96-scal}. In the normal 
state $R=dU/dI$ can be described by Wexler's formula as a sum of Sharvin's 
($R_{\text{SHA}}$) and Maxwell's ($R_{\text{MAX}}$) resistance 
\begin{equation}
        R(T) \approx 2R_{\text{K}}/(ak_{\text{F}})^2 + \rho (T)/2a
\label{equ-1}
\end{equation}
with $R_{\text{K}}=\text{h}/\text{e}^2=25.8\,\text{k}\Omega$, $k_{\text{F}}$ 
the Fermi wave number, and $a$ the contact radius.
For the break junctions normal quasi-particle reflection and the effects of 
a normal interface layer should be negligible. We determine the radius by 
comparing the $T$-dependent part of the resistivity with that of the contact 
resistance. This method works excellently for the HF compounds because of 
their huge $T$-dependence of $\rho (T)=\rho_0+AT^2$ at low $T$ due to the 
strong intrinsic electron-electron interactions. 

Three different mechanism reduce $R$ in the SC state: 
$i)$ The coherent coupling between the SC condensates from both sides of the 
contact can lead to the JE, and $R$ disappears at low current densities. 
$ii)$ Single and multi AR enhance not only the JE supercurrent but also the 
net current at finite bias voltage. 
$iii)$ $R_{\text{MAX}}$ is frozen out. 
While the first two processes are boundary effects that scale with the 
inverse contact area, $R_{\text{MAX}}$ scales with the inverse radius. 

We apply Eq. \ref{equ-1} at zero bias, and make a systematic study on the 
contact size. At $U\approx 0$ local heating and the 
self-magnetic field of the current through the junction can be neglected. 
Thus the contact is at a well-defined condition. And by observing how the 
anomalies vary with radius, one gets an idea about the physics behind them.

Our setup is similar to that described by Muller et al. \cite{Muller92}. The 
samples were cut into $5 - 10\,\text{mm}$ 
long slabs of about $1\times 1\,\text{mm}^2$ cross section, a $\sim 0.5\,$mm
deep nut defines the break position. They were glued electrically isolated 
onto a 0.5 mm thick gold-plated copper-bronze bending beam. A screw, driven 
by two thin (0.3 mm diam.) cotton threads, breaks the sample and makes the 
coarse adjustment. Vertical resolution is about $2\,\mu\text{m}$. A piezo 
tube serves for fine adjustment ($2\,\text{nm}$ resolution). The whole setup 
sits in the vacuum region of the refrigerator. A magnetic field can be 
applied perpendicular to the current flow.

\begin{figure}
\vbox to 220mm{
\epsfysize 215mm
\hskip 20mm\epsffile{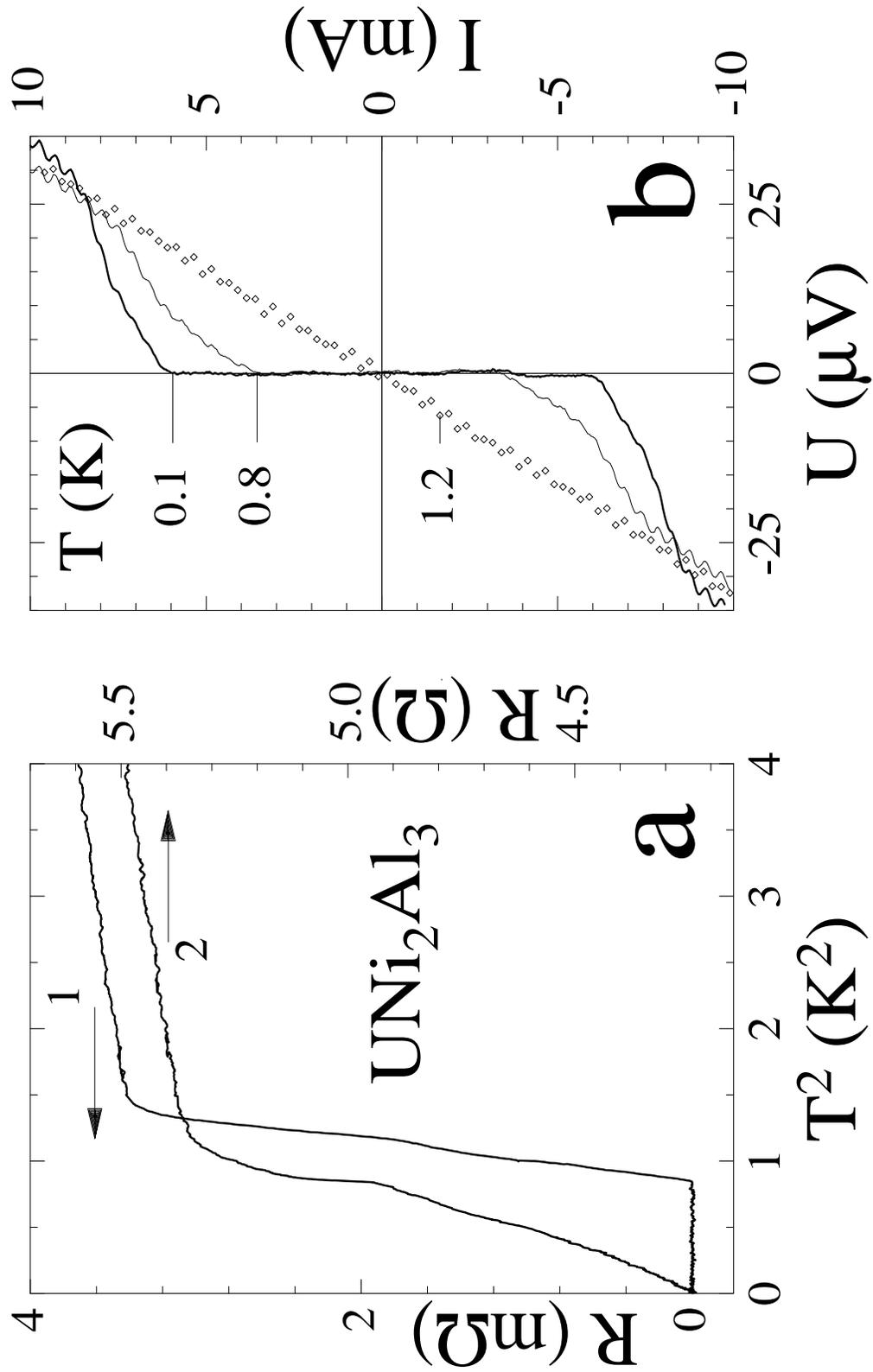}
}
\caption
{a) $R(U=0)$ vs $T^2$ of two different UNi$_2$Al$_3$ junctions. 
 b) $I$ vs $U$ of contact 1 of Fig. a). At low $T$ this contact 
 is probably driven normal due to the contact resistance of the current 
 leads.}
\label{r-vs-t}
\end{figure}

Here we present the results for two UNi$_2$Al$_3$ polycrystals (batch no.
27 100), the other HFSC behave - in most respect - quite similar 
\cite{SCBK96}. Fig. \ref{r-vs-t} (a) shows typical $R(T)$-traces, recorded 
while continuously increasing $T$. The step-like increase $\delta R$ 
indicates the SC transition at $T_{\text{c}}=1.2\,\text{K}$. Comparing 
the slope of $R$ vs. $T^2$ above $T_{\text{c}}$ with 
$A=0.25\,\mu\Omega\text{cm}/\text{K}^2$ yields the radius $a$. On increasing 
the force on the bending beam, the contact area becomes smaller and $R$ 
increases. Finally, the sample breaks and the contact has to be re-adjusted. 
However, the size of the SC signal relative to the slope in the normal 
state remains almost the same, indicating their common origin.

\begin{figure}
\vbox to 220mm{
\epsfysize 215mm
\hskip 20mm\epsffile{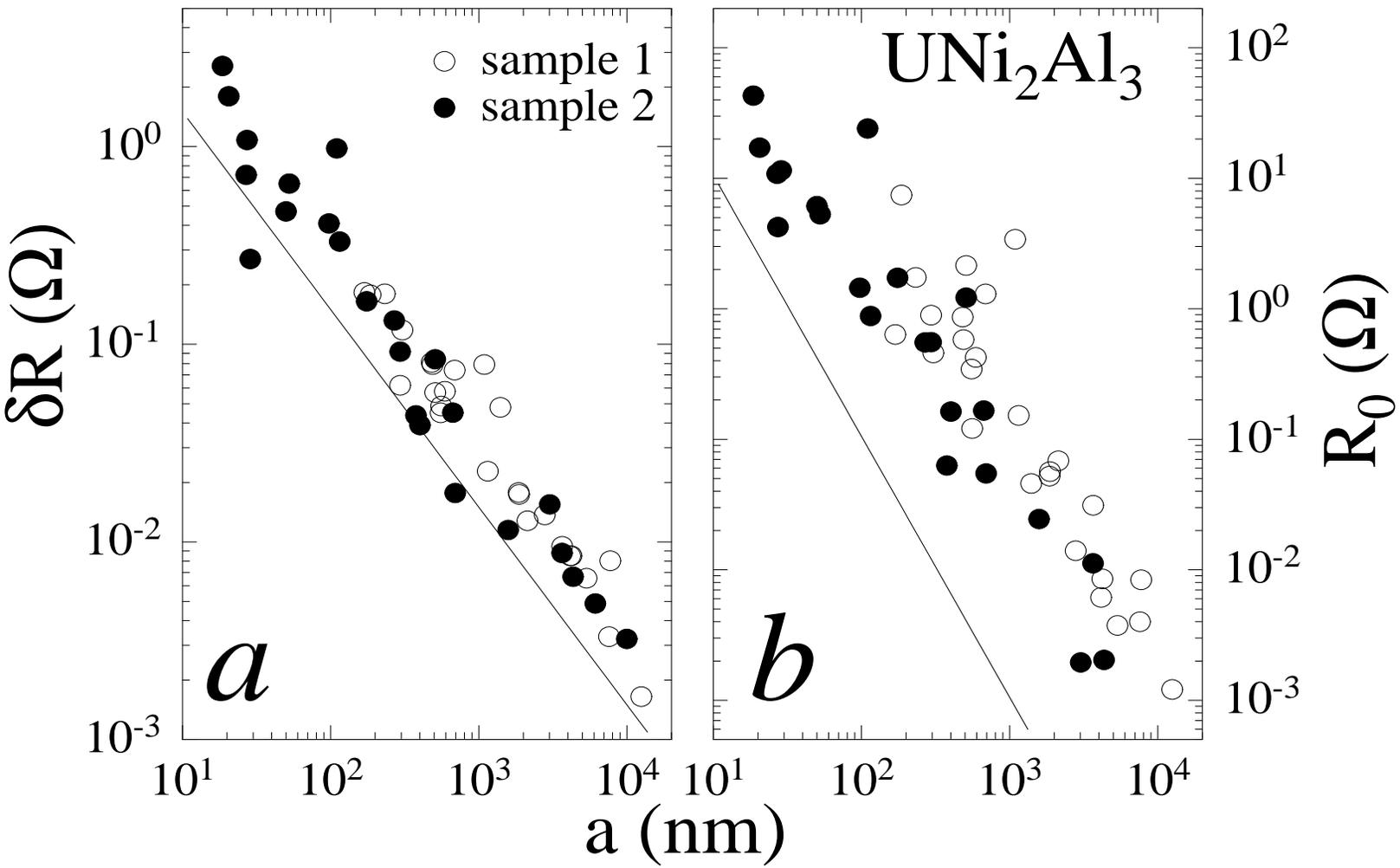}
}
\caption
{a) Size of the SC anomalies $\delta R$ vs radius $a$. 
 The solid line is $R_{\text{MAX}} = \rho_0/2a$. 
 b) Residual resistance $R_0$ vs $a$. The solid line is 
 $R_{\text{SHA}} = 2R_{\text{K}}/(ak_{\text{F}})^2$ at 
 $k_{\text{F}} = 10\,\text{nm}^{-1}$.}
\label{dra}
\end{figure}

Only two out of 50 low-$R$ contacts had unresolvably small residual 
resistances $R_0 = R(T\rightarrow 0) \ll 0.1\,\text{m}\Omega$ like that in 
Fig. \ref{r-vs-t} b). $R_0$ vanishing below $T_{\text{c}}$ might result from 
a supercurrent, but Fig. \ref{dra} (a) shows that 
$\delta R \propto 1/a$ like Maxwell's resistance. Since the average 
$2a\delta R \approx 6\,\mu\Omega\text{cm}$ is near the bulk 
$\rho_0 = 2.9\,\mu\Omega\text{cm}$, contributions from a supercurrent or 
AR are hard to resolve. Note, immediately before breaking the samples had 
$2a\delta R \approx 4\,\text{and}\,6\,\mu\Omega\text{cm}$, respectively, and 
a residual resistance ratio of 25.

$R_0 \propto 1/a^2$ is strongly enhanced with respect to $R_{\text{SHA}}$, 
see Fig. \ref{dra} (b). This corresponds to a rather small effective 
Fermi wave number $k_{\text{F,eff}} \approx 1\,\text{nm}^{-1}$. The 
transition from the ballistic to the thermal regime takes then place at 
$R_0 \approx 0.01\,\Omega$, and a huge elastic mean-free path of 
$l \approx 2\,\mu\text{m}$. 
An alternative explanation for the large $R_0$ is a normal interface layer. 
Direct information about it could be obtained from high-$R$ junctions, 
that have zero-bias maxima (probably from scattering at disordered U 
magnetic moments), but no SC features. 
              
\begin{figure}
\vbox to 220mm{
\epsfysize 215mm
\hskip 20mm\epsffile{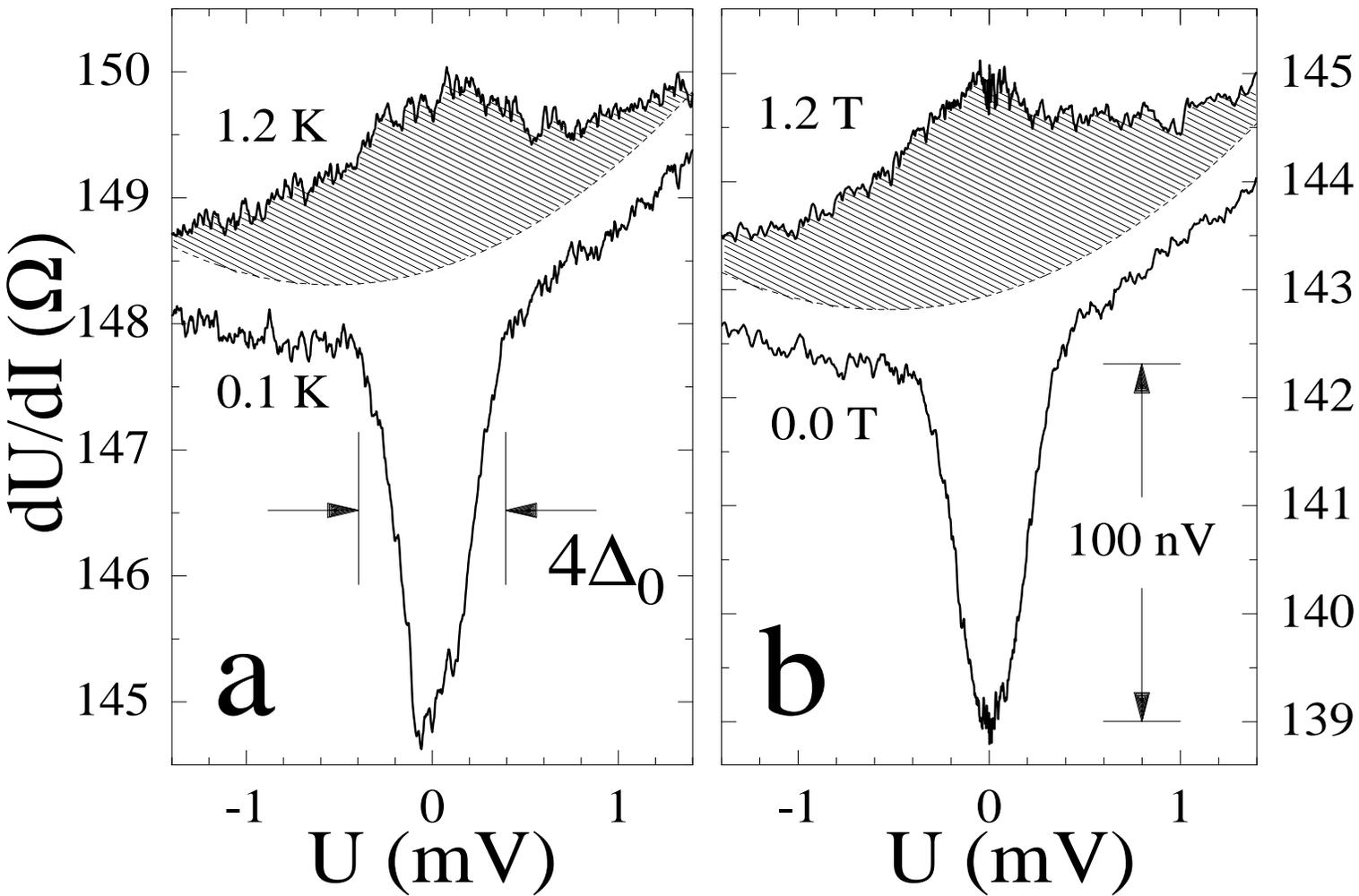}
}
\caption
{$dU/dI$ vs $U$ at a) $B=0\,\text{Tesla}$ and 
b) $T=0.1\,\text{K}$ both in the normal and SC state. 
The hatched area mark tentatively the contribution from the interface layer.
Also indicated is the width of the SC anomaly $4\Delta_0$ as well as the 
signal size at $4\,\mu\text{V}$ excitation.}
\label{spec}
\end{figure}

Fig. \ref{spec} shows spectra of a contact in the intermediate regime with 
both SC features of width $4\Delta_0\approx 0.8\,\text{meV}$,
and zero-bias maxima when the contact is driven normal. From the size of the 
SC anomaly $\delta R \approx 3\,\Omega$ we estimate 
$a \approx 10\,\text{nm}$. Driving the contact normal to above $T_c$
or $B_{c2}$ increases $R$ by
about $0.5\,\Omega$. What we directly 'see' from the 
normal layer is then only $1-2\,\Omega$, i.~e.~a fraction of the total
resistance. This layer must be thinner than the radius to observe the SC 
anomalies. It may add to the pair-breaking effect of the interface and
explain why we could not find SC features by vacuum tunnelling. 

In summary, the break junctions support our previous findings with HF 
heterocontacts. The brickwall to understand these junctions seems to 
be the interface itself. To recover the energy-dependend SC gap from 
the spectra of low-$R$ contacts will require at least detailed knowledge 
of the mean-free path effects described by $R_{\text{MAX}}$.

This work was supported in part by the SFB 252 Darmstadt/Frankfurt/Mainz 
and the BMBF Grant No. 13N6608/1.

\newpage
\onecolumn

\end{document}